\documentclass[a4paper]{article}

\usepackage{icrc2013}
\usepackage[english]{babel}

\title{The 2013 multiwavelength campaign on the Narrow-Line Seyfert 1 galaxy 1H 0323+342: a rosetta stone for the jet/disk paradigm}

\shorttitle{NLSy1s: 1H 0323+342}

\authors{
Omar Tibolla$^{1}$,
Sarah Kaufmann$^{2}$,
Luigi Foschini$^{3}$,
Karl Mannheim$^{1}$,
Shu Zhang$^{4}$,
Jian Li$^{4}$,
Emmanouil Angelakis$^{5}$,
Lars Fuhrmann$^{5}$,
Paul H\"ausner$^{6}$, 
Jannik Kania$^{6}$, 
Dominik Els\"asser$^{1}$,
Annika Kreikenbohm$^{1,7}$,
Robert Schulz$^{1,7}$,
Matthias Kadler$^{1}$.
}

\afiliations{
$^1$ ITPA, Universit\"at W\"urzburg, Campus Hubland Nord, Emil-Fischer-Str. 31 D-97074 W\"urzburg, Germany \\
$^2$ Landessternwarte, Universit\"at Heidelberg, K\"onigstuhl, D 69117 Heidelberg, Germany \\
$^3$ INAF - Osservatorio Astronomico di Brera, Merate (Italy) \\
$^4$ IHEP, Chinese Academy of Science, Beijing, China \\
$^5$ Max Planck Institute f\"ur Radioastronomie, Bonn, Germany \\
$^6$ Friedrich K\"onig Gymnasium, W\"urzburg, Germany \\
$^7$ Dr. Remeis Sternwarte \& ECAP, Universit\"at Erlangen-N\"urnberg, Sternwartstr. 7, 96049, Bamberg, Germany
}

\email{omar.tibolla@gmail.com ; Omar.Tibolla@astro.uni-wuerzburg.de}

\abstract{Narrow-Line Seyfert 1 galaxies have been established as a new class of gamma-ray emitting AGN with relatively low black hole masses, but near-Eddington accretion rates.

Other extragalactic gamma-ray sources observed so far such as Flat Spectrum Radio Quasars, Radio Galaxies, and BL Lacertae Objects generally exhibit much higher black hole masses and, in the case of BL Lac objects and FRI Radio Galaxies, much lower accretions rates. 

The multifrequency campaign of 2013 centered on the bright source 1H 0323+342 will provide further insights into the nature of the jets and their gamma ray production mechanisms in a largely unexplored corner of AGN parameter space. 
Here, we show preliminary results of this campaign and discuss them.}


\begin{document}
\maketitle

\section{Narrow-Line Seyfert 1 galaxies and 1H 0323+342: the disk-jet symbiosis}

According to the standard lore, gamma-ray emission from blazars originates
in jets driven by the spin-down of the ergosphere of a rotating supermassive
black hole \cite{1}. By analogy with the Galactic microquasars \cite{2}, blazars would correspond to states during which the accretion
rate is far below the Eddington limit and the accretion flow forms a torus,
anchoring the currents which provide support for the jet emitting a hard
spectrum. Sub-Eddington accretion rates would also imply that the radiation
field density remains moderate, so as to not quench the acceleration of
particles to high energies.

Narrow-Line Seyfert 1 Galaxies however do not fit into this picture, showing masses of the central black hole smaller than those of blazars and accretion rates close to the Eddington limit (see \cite{fosc} for a review). Particularly, 1H 0323+342 \cite{3} is a source with an optically thick disk \cite{4}, but at the
same time showing gamma-ray emission, as observed with {\it Fermi} \cite{5}. It is important to underline here that the Narrow-Line Seyfert 1 galaxies have typical masses  $< 10^8 M_{sun}$ (while quasars and blazars generally have masses $> 10^8 M_{sun}$) and hence we are really dealing with a new class of gamma-ray emitting active galactic nuclei (AGN).  There are even some clues that 1H 0323+342 might show TeV radiation \cite{6} and, moreover, it is the  closest (z=0.061) the among the GeV gamma-ray emitting NLSy1s.
Given that the location of the gamma-ray emitting zone in other extragalactic
radio sources has been found to be as close as $100 R_G$ to the central
black hole  \cite{7}, this would be at odds with the
high radiation density (and thus pair production
probability) of the central parsec in Eddington accreting Seyferts.

Since the source does not fit into the extremal hard or soft states, it
might, in fact, represent a transitional case in which the inner disk heats
up and blows up to a torus configuration, as the accretion rate goes down
(or vice versa). 

The Galactic X-ray binaries are known to switch between these two states repeatedly due to disk instabilities, which for AGN disks would have typical time scales of $10^4$ - $10^5$ years
\cite{8} and irradiation of the outer disk by the central
hot torus.  Given the characteristic time scale of Seyfert 1 activity of
$10^8$ yrs, the switching should result in about $10^{-4}$ - $10^{-3}$ of
the AGN to be in an intermediate state, which is quite in line with the low
number of gamma-ray emitting Narrow-Line Seyfert 1 galaxys that {\it Fermi} has detected so far.

However, to have simultaneously a jet and a high accretion rate, the problem of how the gamma rays and the hard X rays (up to MeV energies) can survive from passing through the condensed X-ray field still remains unresolved. Further clues  to this direction may come from the investigations on the flux variability and on the possible spin of the BH. Since it is known that the high energy emission is variable, the jet may be formed accompanying with relatively weak soft X-ray, as was commonly seen in XRB.  A simultaneous monitoring on source at gamma-rays by {\it Fermi}, at hard X-rays by {\it Integral} and at softer X-rays by {\it Swift} will make it  possible to investigate the correlation and time lag between X-rays and gamma-rays. In case that a time lag or an anti-correlation is derived, a scenario of having a temporarily formed jet under relatively low X-ray field will be enforced. 

An alternative would be to have the contemporary jet for gamma rays and high accretion rate for X rays. In this case, the dilution due to interaction between gamma rays and X rays can be effectively avoided by introducing into the system a highly spinning BH.  One example to this direction can be found in \cite{9}, where the spin of M87 can be constrained by gamma-ray observations. The transparent radii to TeV gamma-rays can change from 5 Rs to 14 Rs for a highly spinning and none-spinning BH. 

The fact that we see a powerful radio (and gamma-ray) jet implies that the
last marginal orbit lies much closer to the black hole than in the
Schwarzschild geometry, if the jet is indeed produced by the
Blandford-Znajek or similar mechanism.  

As can be seen in Figure 1, the spectral energy distribution  (SED) of NLSy1 glaxies is rather complex. In the energy range of {\it Integral} from $4$ keV to $10$ MeV, the dominant emission process is the Synchrotron-Self-Compton (SSC) emission from the high energetic jet and the External Compton (EC) emission in which the synchrotron photons interact with a photon field close to the jet. The EC component is necessary to describe the detected GeV $\gamma$-ray emission. Constraints on the model can be obtained with {\it Integral} observations on NLSy1 galaxies which will provide a good coverage of the gap between the soft X-ray and the GeV $\gamma$-ray emission.

Moreover, since in the recent years it has been debated whether NLSy1 galaxies are indeed a new class of AGN (e.g. \cite{5}) or they are normal flat-spectrum radio quasars (FSRQ; e.g. \cite{dam}), there is another very important feature that NLSy1 galaxies have in common with this other class of AGN and that must be carefully investigated:the MeV peaked spectral emission.
In fact there are two types of FSRQs: those with steep gamma-ray spectrum (MeV blazars) and those with flat gamma-ray spectrum (GeV blazars). 

The two types of blazars are thought to be two aspects of the same phenomenon (e.g. \cite{13}) and it has been predicted that:

\begin{itemize}
\item variability timescales of MeV blazars should be longer than variability timescales of GeV blazars;
\item both types of blazars are expected to appear in the same object.
\end{itemize}

Simultaneous observations at GeV and MeV energies will be able to answer these questions.

Moreover, \cite{14} and \cite{15} point out that we can explain only $\sim$16\% of extragalactic MeV-GeV background and consequently need for a numerous faint population of AGN $>$ 10 MeV in order to be able to explain the observed isotropic extragalactic gamma-ray diffuse emission.

The analogy with MeV blazars, poorly known due to the so-called MeV-gap, can be succesfully investigated by {\it Integral} from the lower energy side.
As shown in \cite{14}, most of the blazars show a general spectral trend of peaking at MeV energies (with several impliactions mentioned above). It will be crucial to see whether NLSy1s fall into this category as well, as suggested/modeled in \cite{5}, and hence to confirm that NLSy1 galaxies are indeed a new class of AGN.

The NLSy1 galaxies with detected GeV emission show very bright peaks in the spectral energy distribution described as EC emission. Therefore the emission expected for the {\it Integral} energy range is rather bright. 


 \begin{figure}[t]
  \centering
  \includegraphics[width=0.45\textwidth]{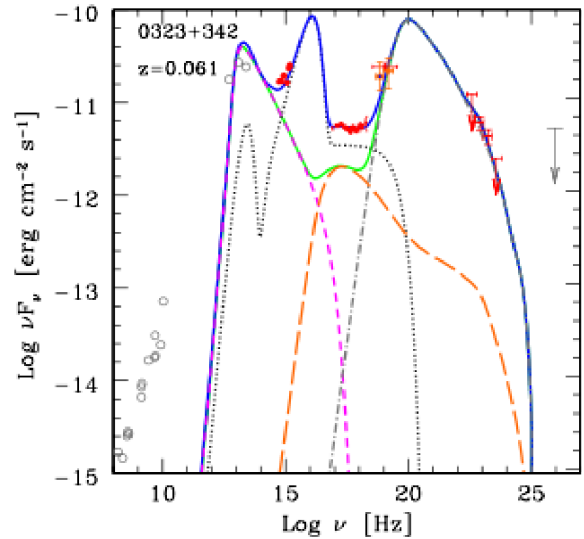}
  \caption{Spectral energy distribution of the NLSy1 galaxy 1H0323+342 from \cite{5}.}
  \label{0323SED}
 \end{figure}

\section{1H 0323+342: precedent campaigns} 

During {\it Swift} observations of 1H0323+342, an indication for a Fe K-alpha line was detected, supporting the model described \cite{5}.
Within these observations, also spectral variability  was detected in the observed X-ray energy band. In general the photon index was soft, as in a typical NLSy1 galaxy, but sometimes it displays a hard tail \cite{12}. 
Moreover this source was observed several times in the pre-{\it Fermi} era. Currently, there are 64 {\it Swift} observations and one {\it Suzaku} (public) plus one accepted target in the last {\it Suzaku} AO (both by {\it Fermi}-LAT collaboration).

Among the four gamma-ray bright NLSy1 galaxies 1H 0323+342 is the brightest at {\it Integral} energies and in fact it has already been detected with {\it Integral} and {\it Swift}/BAT \cite{12}. This is crucial for variability studies, since its brightness allows us to observe 1H0323+342 with short observations. 

Moreover, 1H0323+342  is the best target among the four gamma-ray bright NLSy1 galaxies also thanks to its high variability. In fact, both {\it Integral} and {\it Swift} \cite{12} observations showed a certain flux and spectral variability, e.g. in 2004 during a 200 ks {\it Integral} observation the source was detected with a low flux and a soft spectrum (2.5 mcrab in the 20-40 keV band; no detection and upper limit determined at 2.6 mcrab in the 40-100 keV band).
Instead, integrating 50 ks of {\it Swift} on-axis observations performed between 2006 and 2008, the source was in a high flux state and with a very hard spectrum (no detection and upper limit determined at 20 mcrab in the 20-40 keV band; 16 mcrab in the 40-100 keV band).

This is in good agreement with what {\it Swift}/XRT and UVOT observed. Under the condition of low optical and X-ray flux the source has a flat spectrum with photon index of around 2. When the optical/UV flux increased (this is generally a surprising feature for a NLSy1 galaxy \cite{16}) also the X-ray flux increased but overall a break around 3 keV appeared with a hard tail of photon index of around 1.4. 
This seems to be perfectly in line with the transitional case in which the inner disk heats up and blows up to a torus configuration, described in the previous section; an alternative to this scenario could be represented by the saturated comptonization \cite{17}: i.e. with the appearence of the jet, the torus is more effective in comptonizing.


 \begin{figure}[t]
  \centering
  \includegraphics[width=0.45\textwidth]{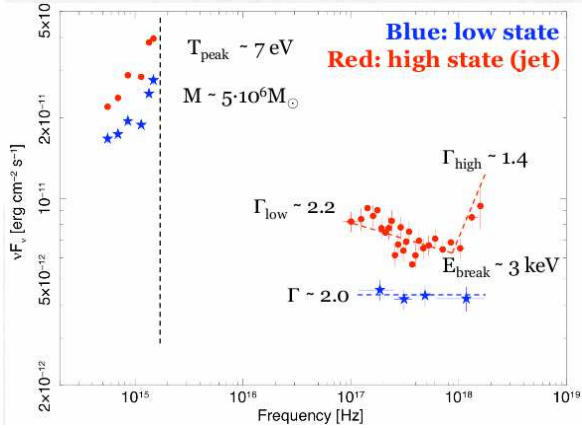}
  \caption{From \cite{fosc}. The {\it Swift}/XRT and UVOT flux distribution for the low and high flux state as described in the text which show the flux and spectral variability in the low energy band. Similar behaviour has been seen in the hard X-ray observations by {\it Integral} and {\it Swift} during this epochs \cite{12}.}
  \label{0323Swift}
 \end{figure}

\section{1H 0323+342: the 2013 deep campaign}

In order to study in details the emission processes and especially the jet-disk coupling in 1H 0323+342,  we established a new multi-wavelength (MWL) campaign with simultaneous observations in the UV/X-ray/hard X-ray bands:

\begin{itemize}
\item 600 ks of observations with {\it Integral} has been accepted (grade A; PI: O. Tibolla) in 2013, divided in four observations of 150 ks each.
\item Sided by {\it Swift} snapshot observations (ToO).
\end{itemize}

In addition, in order to have multi-frequency spectra from this enigmatic source ranging from radio to gamma-rays, we observe the source also:

\begin{itemize}
\item in GeV gamma rays with {\it Fermi}-LAT;
\item in optical with the 50 cm Hans-Haffner-Telescope (Hettstadt, W\"urzburg);
\item at radio frequencies, by means of a dedicated high-sensitivity, multi-frequency cm/mm radio spectral monitoring with the Effelsberg 100 m telescope, as part of the large F-Gamma program\footnote{http://www.mpifr-bonn.mpg.de/div/vlbi/fgamma/fgamma.html} \cite{fgamma} \cite{fgamma2} every few weeks at 2.6, 4.8, 8.3, 10.5, 15, 23, 32 and 43 GHz.
\item at radio frequencies, structural changes of the parsec-scale jets will be monitored as part of the VLBI monitoring program MOJAVE\footnote{http://www.physics.purdue.edu/astro/mojave} \cite{mojave}.
\end{itemize}

The first two {\it Integral} 150 ks observations were succesfully taken so far from January 13 - 15 and February 15 - 16, 2013. 
The other MWL facilities observed 1H 0323+342 simultaneously to this time ranges.
The campaign is still ongoing and all presented results are preliminary.

With all {\it Integral} data combined, 1H 0323+342 is detected in the energy range 20-40 keV with a significance of $\sim 3$ standard deviations in 54.2 ks effective exposure; the average flux is 0.25$\pm$0.08 counts/s. The source is not detected in the energy range 40-100 keV, with a significance of 1.6 standard deviations in 85 ks effective exposure; the average flux is 0.12$\pm$0.08 counts/s.

With {\it Integral} we have 141 Science Windows (ScW) so far, but only 70 of them have good time intervals which are long enough to produce a sky map.
During most of ScWs, 1H 0323+342's significance is below 3 standard deviations; only two ScWs have significance larger than 3 standard deviations but still below 4 standard deviations. One of them have a high flux (28.98 counts/s, near MJD 56338) but large error bar. The effective exposure of this Science Window is relatively short $\sim$170 s. It is not clear whether there was a flare or not.

Even if from {\it Integral} observations it is not clear if 1H 0323+342 is flaring in February or not, two different states for the source seem detected with {\it Swift}: in fact if we compare the data of January the 13$^{th}$ (Fig. 3) with the data of February  the 15$^{th}$ (Fig. 4), we note a very similar behavior to the one described in Fig. 2.

 \begin{figure}[t]
  \centering
  \includegraphics[width=0.45\textwidth]{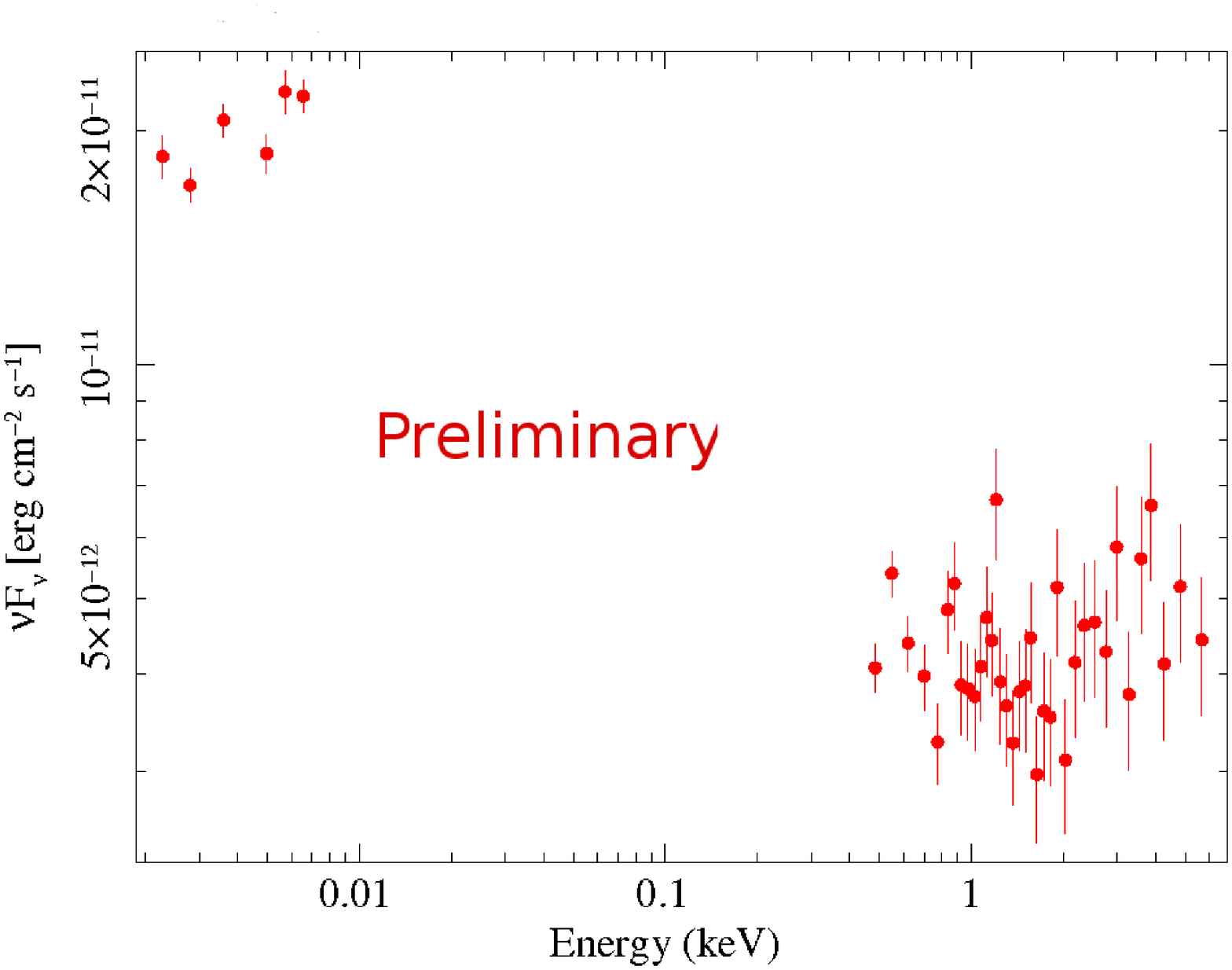}
  \caption{ {\it Swift} observations of 1H 0323+342, taken the 13$^{th}$ of January 2013; a flat X-ray spectrum with $\Gamma \simeq 2$ is the best description for the observations.}
  \label{Swift1}
 \end{figure}

 \begin{figure}[t]
  \centering
  \includegraphics[width=0.45\textwidth]{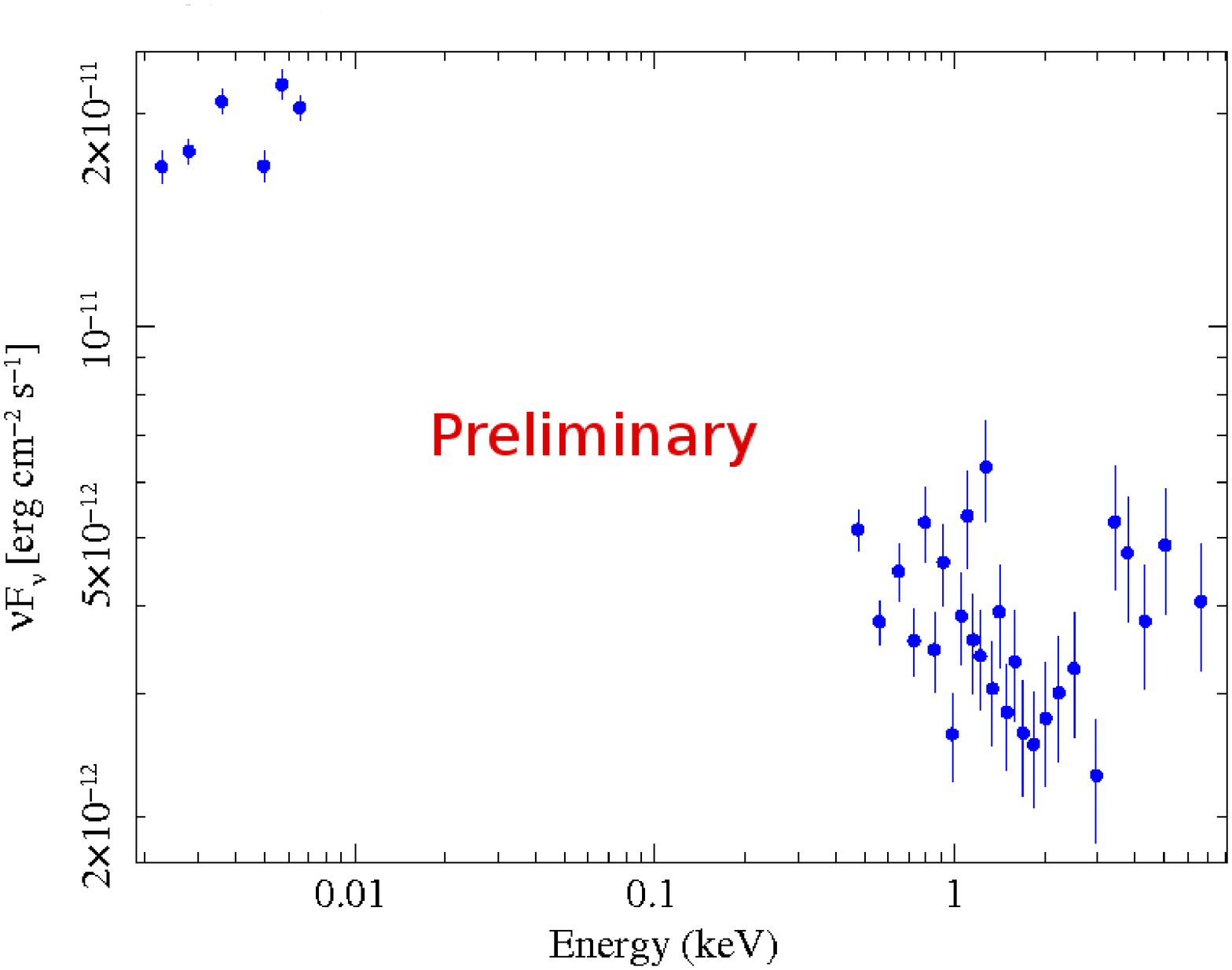}
  \caption{ {\it Swift} observations of 1H 0323+342, taken the 15$^{th}$ of February 2013; the X-ray spectrum is well described by a broken power-low with $\Gamma_1 \simeq 2.3$, $\Gamma_2 \simeq 1.6$ and $E_b \simeq 2$ keV.}
  \label{Swift2}
 \end{figure}

The UV and optical observations by {\it Swift}/UVOT shown in Figure 3 and 4 have been corrected for Galactic extinction using\footnote{http://irsa.ipac.caltech.edu/applications/DUST/} $E(B-V) = 0.2095 \; \rm{mag}$  and the X-ray spectra have been corrected for the Galactic absorption of $N_{\rm{H}} = 1.27 \times 10^{21} \; \rm{cm^{-2}}$ \cite{18}. 

In fact, a flat X-ray spectrum is the best description for the observations of January the 13$^{th}$ (Fig. 3), with spectral index $\Gamma \simeq 2$; while a broken power-low slope with a hard tail is the best description for the observations of February the 15$^{th}$ (Fig. 4): the two spectral index are $\Gamma_1 \simeq 2.3$ and $\Gamma_2 \simeq 1.6$ with a break at $\sim$2 keV (slightly different from the high state values shown in Fig.2). 

In optical, observations with the 50cm Hans-Haffner-Telescope revealed a brightness in the 
Bessel R-band of $15.49\pm 0.02 \; \rm{mag}$ in January and $15.30 \pm 0.02 \; \rm{mag}$ in February, 2013.
Therefore, an increase in optical brightness was detected contemporaneously to the second {\it Integral} observations. 

In the long term monitoring with Effelsberg, 1H 0323+342  show a rather constant flux of $\sim 0.5 \; \rm{Jy}$ in 2010 and 2011. At the beginning of 2013, a radio flare appeared with a  flux raised up to $\sim 1.5 \; \rm{Jy}$ \cite{19}. At the time contemporaneous to the {\it Integral} observations, the spectral shape changed slightly.

Even if the results of the deep MWL on-going campaign on 1H 0323+342 in 2013 are still preliminary, we can confirm that the variation of fluxes and of spectra give the direct view to the changes in the disk emission.


\end{document}